\documentclass[prl,nofootinbib,nobibnotes,twocolumn,superscriptaddress]{revtex4-1}

\usepackage{graphicx}
\usepackage{amsmath,amsfonts,amsthm,amssymb} 
\usepackage[caption=false]{subfig}
\usepackage{amsmath}
\usepackage{mathrsfs}
\usepackage{dsfont}
\usepackage{float}
\usepackage{siunitx}
\usepackage{adjustbox}
\usepackage{color}
\usepackage{bbold}
\usepackage{lipsum}
 
\usepackage[utf8]{inputenc}
\usepackage{times}
\usepackage{upgreek}
\usepackage{physics}
\usepackage{bm}
\usepackage{mathtools}% http://ctan.org/pkg/mathtools

\begin{document}

\title{Tunable pseudo-magnetic fields for polaritons in strained metasurfaces}

\author{Charlie-Ray Mann}
% \email{cm433@exeter.ac.uk}
\affiliation{School of Physics and Astronomy, University of Exeter, Exeter, EX4 4QL, United Kingdom.}

\author{Simon A. R. Horsley}
\affiliation{School of Physics and Astronomy, University of Exeter, Exeter, EX4 4QL, United Kingdom.}

\author{Eros Mariani}
\affiliation{School of Physics and Astronomy, University of Exeter, Exeter, EX4 4QL, United Kingdom.}

\begin{abstract}
\noindent Artificial magnetic fields are revolutionizing our ability to manipulate neutral particles, by enabling the emulation of exotic phenomena once thought to be exclusive to charged particles \cite{Hafezi2011,Umucalilar2011,Fang2012a,Fang2012b,Fang2013b,Hafezi2013,Lin2014,Tzuang2014,Liu2015,Schine2016,Schomerus2013,Rechtsman2013,Abbaszadeh2017,Brendel2017,Yang2017,Wen2019,Jia2019,Peri2019}. In particular, pseudo-magnetic fields generated by nonuniform strain in artificial lattices have attracted considerable interest because of their simple geometrical origin \cite{Schomerus2013,Rechtsman2013,Jia2019,Yang2017,Wen2019,Abbaszadeh2017,Brendel2017,Peri2019}. However, to date, these strain-induced pseudo-magnetic fields have failed to emulate the tunability of real magnetic fields because they are dictated solely by the strain configuration. Here, we overcome this apparent limitation for polaritons supported by strained metasurfaces, which can be realized with classical dipole antennas or quantum dipole emitters. Without altering the strain configuration, we unveil how one can tune the pseudo-magnetic field by modifying the electromagnetic environment via an enclosing photonic cavity which modifies the nature of the interactions between the dipoles. Remarkably, due to the competition between short-range Coulomb interactions and long-range photon-mediated interactions, we find that the pseudo-magnetic field can be entirely switched off at a critical cavity height for any strain configuration. Consequently, by varying only the cavity height, we demonstrate a tunable Lorentz-like force that can be switched on/off and an unprecedented collapse and revival of polariton Landau levels. Unlocking this tunable pseudo-magnetism for the first time poses new intriguing questions beyond the paradigm of conventional tight-binding physics.
\end{abstract}

\maketitle

\noindent Unfortunately, neutral particles do not directly couple to the electromagnetic gauge potentials. Therefore, exotic phenomena exhibited by charged particles in magnetic fields, such as the Lorentz force, Aharonov-Bohm effect, and Landau quantization, remain elusive for neutral particles. However, it was recently demonstrated that nonuniform strain in graphene can generate pseudo-magnetic fields which can mimic some of the properties of real ones \cite{Guinea2009,Levy2010,Low2010,deJuan2011}.  This tantalizing prospect inspired many attempts to emulate strain-induced pseudo-magnetic fields for neutral particles in acoustics \cite{Yang2017,Wen2019}, phononics \cite{Abbaszadeh2017,Brendel2017}, and photonics \cite{Schomerus2013,Rechtsman2013}, by judiciously engineering aperiodicity in artificial lattices that mimic the tight-binding physics of graphene. However, these strain-induced pseudo-magnetic fields have failed to emulate one key property of real magnetic fields: \emph{tunability}. While real magnetic fields can be tuned by varying external parameters in the lab, these emergent pseudo-magnetic fields are dictated solely by the strain configuration, rendering them fixed by design. Therefore, a fundamental question arises:  can we overcome this seemingly intrinsic lack of tunability?

Here we show that tunable pseudo-magnetic fields for neutral particles are indeed possible by going beyond the paradigm of conventional tight-binding physics. Specifically, we consider polaritons supported by a strained honeycomb metasurface composed of resonant dipole scatterers. This simple model can be realized in a variety of experimental platforms, ranging from microwave \cite{Mann2018}, plasmonic \cite{Mann2018,Weick2013, Lamowski2018}, or dielectric \cite{Slobozhanyuk2015} metamaterials composed of classical dipole antennas, to arrays of excitonic nanostructures \cite{Yuen-Zhou2016} or atom-like quantum emitters \cite{Perczel2017}. Crucially, the resonant interaction with light results in long-range interactions between the dipoles which qualitatively depend on the surrounding electromagnetic environment. Consequently, previous results derived from conventional tight-binding models do not trivially extend to lattices of interacting dipoles \cite{Mann2018}. By exploiting this key difference, we show that without altering the strain configuration one can tune the \emph{artificial} magnetic field by modifying the \emph{real} electromagnetic environment.

To demonstrate this tunability we embed the metasurface inside a planar photonic cavity, where one can modify the nature of the long-range dipole-dipole interactions by simply varying the cavity height. As a result, one can tune the strength of the pseudo-magnetic field acting on the polaritons and even switch it off entirely at a critical cavity height, despite the strain configuration being fixed --- this highly non-trivial result is impossible to achieve with conventional tight-binding systems. Consequently, we demonstrate a tunable Lorentz-like force that can be switched on and off, deflecting polariton wavepackets into effective cyclotron orbits whose radius can be controlled via the cavity height. For large strains, we also demonstrate Landau quantization for the polaritons where progressively decreasing the cavity height can induce an unprecedented collapse and revival of the polariton Landau levels. 

%%TC:ignore
\begin{figure*}[t]
\includegraphics[width=0.8\textwidth]{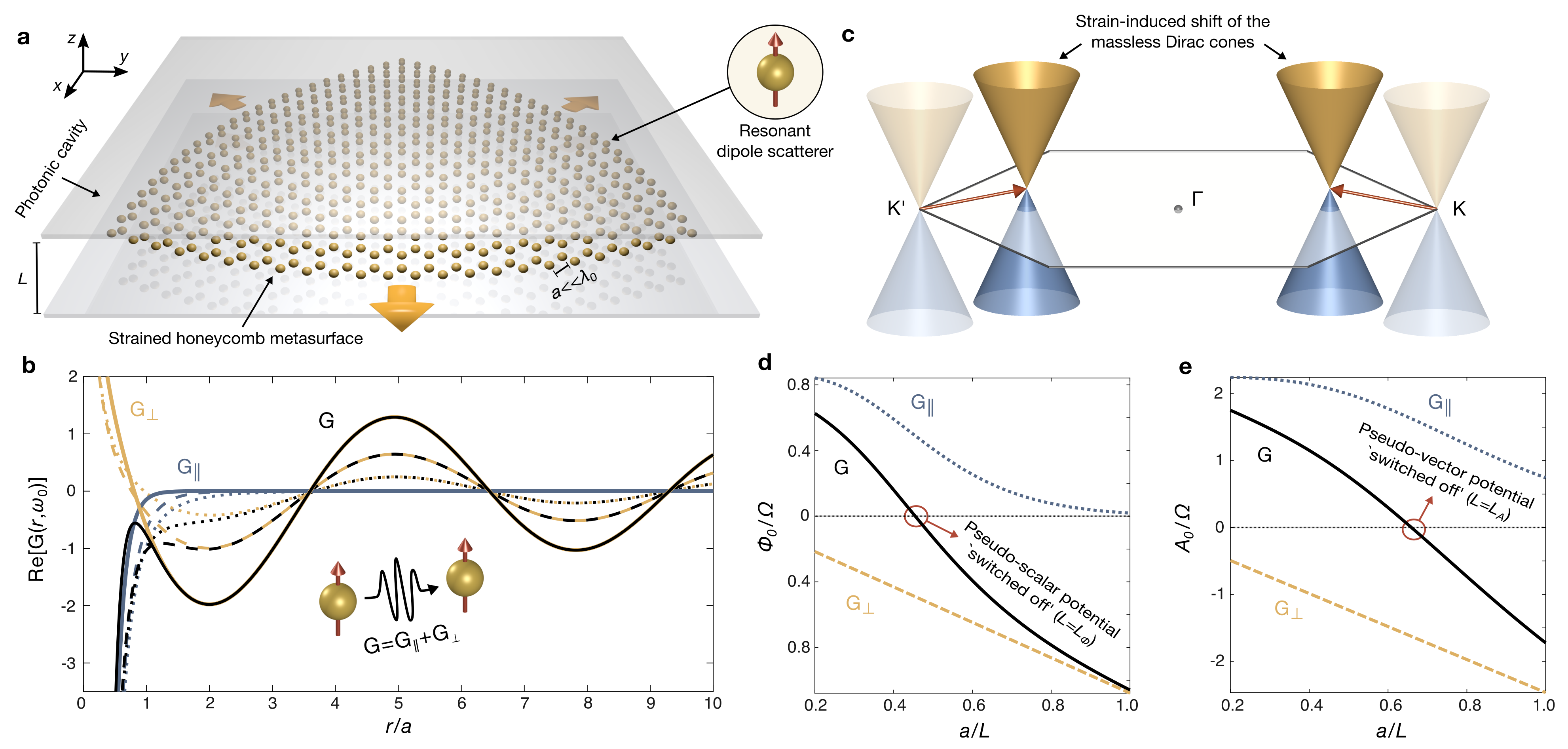}
\caption{\textbf{Cavity-tunable pseudo-gauge potentials for polaritons in strained metasurfaces.} \textbf{a} Schematic of a trigonally strained honeycomb metasurface composed of two inequivalent hexagonal sublattices of resonant dipole scatterers (e.g., classical antennas or quantum emitters), which is embedded inside a planar photonic cavity with height $L$. We assume that the scatterers have an anisotropic polarizability such that the induced dipole moments point in the $\hat{\mathbf{z}}$ direction (see inset). The unstrained metasurface has a subwavelength nearest-neighbor separation $a\ll\lambda_0$ ($\lambda_0=2\pi c/\omega_0$) and therefore supports evanescently bound surface polaritons. \textbf{b} Shows how the real part of the cavity Green's function $\mathsf{G}$ (evaluated at the free-space resonant frequency $\omega_0$) varies with the distance $r$ from the source (black lines) for cavity heights of $a/L=0.2$ (dotted lines), $a/L=0.5$ (dashed lines), and $a/L=1$ (solid lines). We also show the separate longitudinal $\mathsf{G}_{\parallel}$ (blue lines) and transverse $\mathsf{G}_{\perp}$ (orange lines) components which describe the short-range Coulomb interactions and long-range photon-mediated interactions, respectively. The unstrained metasurface supports massless Dirac polaritons characterized by linear Dirac cones near the high symmetry $\mathrm{K}$/$\mathrm{K}'$ points located at $\pm\mathbf{K}=\pm(4\pi/3\sqrt{3}a,0)$ in the first Brillouin zone as depicted in \textbf{c}. The applied strain leads to a shift of the Dirac cones in frequency and momentum which is described by a pseudo-scalar and pseudo-vector potential, respectively. \textbf{d}-\textbf{e} Show how the strain-independent parameters in the pseudo-scalar potential ($\Phi_0$) and pseudo-vector potential ($A_0$) vary as a function of the cavity height (black solid lines), where we also show the separate contributions emerging from the Coulomb interactions (blue dotted lines) and the photon-mediated interactions (orange dashed lines). At critical cavity heights ($L_{\Phi}$ and $L_{A}$) there is perfect cancellation between these contributions resulting in the pseudo-gauge potentials being switched off for any strain configuration. Plots obtained with parameters $\omega_0=c|\mathbf{K}|/2.2$ and $\Omega=0.01\omega_0$.}
\label{fig:tunablegaugefields}
\end{figure*}
%%TC:endignore

We model the resonant dipole scatterers with a generic bare polarizability $\alpha_0(\omega)=2\omega_0\Omega(\omega_0^2-\omega^2-\mathrm{i}\omega\Gamma)^{-1}$ that is applicable to both classical antennas and quantum emitters in their linear regime, where $\Omega$ characterizes the strength of the polarizability, $\Gamma$ accounts for non-radiative losses, and $\omega_0$ is the free-space resonant frequency. The metasurface is formed by arranging the scatterers into a honeycomb array consisting of two inequivalent hexagonal sublattices. We assume the scatterers to be anisotropic such that the induced dipole moments point normal to the plane of the lattice. Furthermore, we consider subwavelength lattice spacing so that the metasurface supports evanescently bound surface polaritons. Finally, the metasurface is embedded at the center of a planar photonic cavity of height $L$, as schematically depicted in figure\,\ref{fig:tunablegaugefields}a, where the cavity walls are assumed to be perfectly reflecting mirrors. 

The interactions between the dipoles are mediated by the cavity Green's function which can be decomposed into its longitudinal ($\parallel$) and transverse ($\perp$) components $\mathsf{G}=\mathsf{G}_{\parallel}+\mathsf{G}_{\perp}$, where $\mathsf{G}_{\parallel}$ describes short-range Coulomb interactions and $\mathsf{G}_{\perp}$ describes long-range interactions mediated by the cavity photons (see Methods for the expressions). In figure\,\ref{fig:tunablegaugefields}b, we show how one can modify the Green's function by varying the cavity height. Naively, it is tempting to assume that the near-field Coulomb interactions dominate the physics given the subwavelength spacing of the metasurface. However, as the cavity height is reduced the Coulomb interactions (blue lines) are exponentially suppressed and the photon-mediated interactions (orange lines) become increasingly dominant, even between nearest-neighbors.

%%TC:ignore
\begin{figure*}[t]
\includegraphics[width=0.8\textwidth]{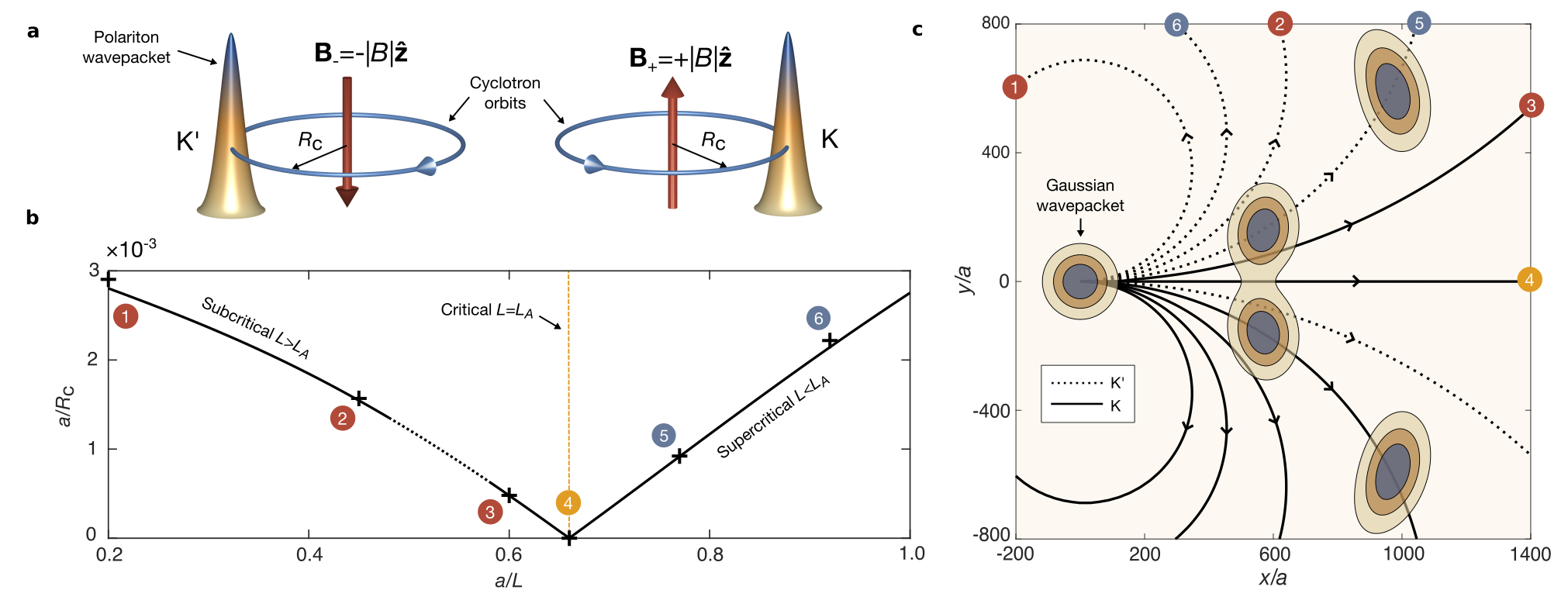}
\caption{\textbf{Cavity-tunable Lorentz-like force and cyclotron motion of polariton wavepackets.} \textbf{a} Schematic depiction of the cyclotron motion exhibited by polariton wavepackets due to a Lorentz-like force generated by a pseudo-magnetic field which has opposite signs in the $\mathrm{K}$/$\mathrm{K}'$ valleys. \textbf{b} Shows how the cyclotron orbit radius varies with cavity height for a fixed frequency relative to the Dirac point and  fixed strain magnitude. The dotted line indicates the region of cavity heights where the linear Dirac cone approximation breaks down (see Supplementary Information for more details). \textbf{c} Centre-of-mass trajectories of Gaussian wavepackets propagating in a trigonally strained metasurface with fixed strain magnitude, but different cavity heights (see Methods for details). Wavepackets in the $\mathrm{K}$ (solid lines) and $\mathrm{K}'$ (dotted lines) valleys undergo cyclotron motion in opposite directions (e.g., see snapshots along trajectory 5), and the calculated radii match very well with the analytical predictions (see crosses in \textbf{b}). For subcritical cavity heights $L>L_{A}$, the cyclotron radius expands as the cavity height is reduced (trajectories 1-3), until a critical height $L=L_{A}$, where the pseudo-magnetic field is switched off and the wavepackets feel no Lorentz-like force (trajectory 4). For supercritical cavity heights $L<L_{A}$, the cyclotron orbits reemerge and the orbit radius shrinks as the cavity height is reduced further (trajectories 5-6). Plots obtained with parameters $\Delta=2\times10^{-5}, \omega_0=c|\mathbf{K}|/2.2$, $\Omega=0.01\omega_0$, and $\delta\omega=-0.001\omega_0$.}
\label{fig:tunableorbits}
\end{figure*}
%%TC:endignore

To gain analytical insight, we have used a coupled-dipole model to derive an effective Hamiltonian describing the polaritons near the $\mathrm{K}/\mathrm{K}'$ points, which is valid to leading order in the strain tensor $\varepsilon_{ij}(\mathbf{r})=(\partial u_j / \partial r_i+\partial u_i / \partial r_j)/2$ where $\mathbf{u}(\mathbf{r})$ is the slowly-varying displacement field. For the $\mathrm{K}$ valley, the effective Hamiltonian reads ($\hbar=1$)
%-------------------------------------
\begin{equation}
\mathcal{H}_{\mathrm{K}} = \omega_{\text{D}}^{}(L)\mathbb{1}+\mathrm{i} v_{\text{D}}(L) \boldsymbol{\sigma}\cdot \grad+\Phi(\mathbf{r},L)\mathbb{1}+ \boldsymbol{\sigma}\cdot \mathbf{A}_{}(\mathbf{r},L)\,,
\label{eq:EffectHamiltonianK}
\end{equation}
%-------------------------------------
\noindent where $\mathbb{1}$ is the identity matrix and $\boldsymbol{\sigma}=(\sigma_x,\sigma_y)$ is the vector of Pauli matrices acting in the sublattice space (see Methods for the derivation and equivalent Hamiltonian for the $\mathrm{K}'$ valley). Therefore, the polaritons behave like massless Dirac quasiparticles with a linear Dirac cone dispersion \cite{Mann2018}, where the Dirac frequency $\omega_{\text{D}}^{}(L)$ and the Dirac velocity $v_{\text{D}}(L)$ can be tuned by varying the cavity height (see Methods for the expressions). As schematically shown in figure\,\ref{fig:tunablegaugefields}c, the strain leads to a spatially-varying shift of the Dirac cone in frequency and momentum which is effectively described by a pseudo-scalar potential $\Phi(\mathbf{r},L)=\Phi_0(L)[\mathsf{\varepsilon}_{xx}(\mathbf{r})+\mathsf{\varepsilon}_{yy}(\mathbf{r})]$, and a pseudo-vector potential $\mathbf{A}_{ }(\mathbf{r},L)= A_0(L)[\varepsilon_{xx}(\mathbf{r})-\varepsilon_{yy}(\mathbf{r}),-2\varepsilon_{xy}(\mathbf{r})]$, respectively. In figures\,\ref{fig:tunablegaugefields}d-e we show how the strain-independent parameters $\Phi_0(L)$ and $A_0(L)$ can be tuned by varying only the cavity height (see Methods for the expressions). Remarkably, there exists critical cavity heights ($L_{\Phi}$ and $L_{A}$) where these parameters vanish identically, thereby switching off the pseudo-gauge potentials entirely for any strain configuration.

To elucidate the physical origin of the vanishing pseudo-gauge potentials, one can notice that the pseudo-vector (scalar) potential is equal to the change in intersublattice (intrasublattice) interaction energy induced by the strain at the high-symmetry points. Interestingly, the change in energy arising from the Coulomb interactions and photon-mediated interactions have opposite signs and thus tend to compensate each other in the pseudo-gauge potentials (see dotted and dashed lines, respectively, in figure\,\ref{fig:tunablegaugefields}d-e).  At the critical cavity heights, the change in Coulomb interaction energy is perfectly cancelled by the change in photon-mediated interaction energy, resulting in no variation of the \emph{total} interaction energy, despite the distances between the dipoles being modified. We wish to emphasize the novelty of this phenomenon: within a nearest-neighbor tight-binding model, a vanishing pseudo-vector potential would demand a hopping parameter that does not vary with distance, which is impossible to achieve with graphene and its artificial analogs. Therefore, this ability to switch off and tune the magnitude of the pseudo-gauge potentials without altering the lattice structure opens up new perspectives beyond conventional tight-binding models.

%%TC:ignore
\begin{figure*}[t]
\includegraphics[width=0.85\textwidth]{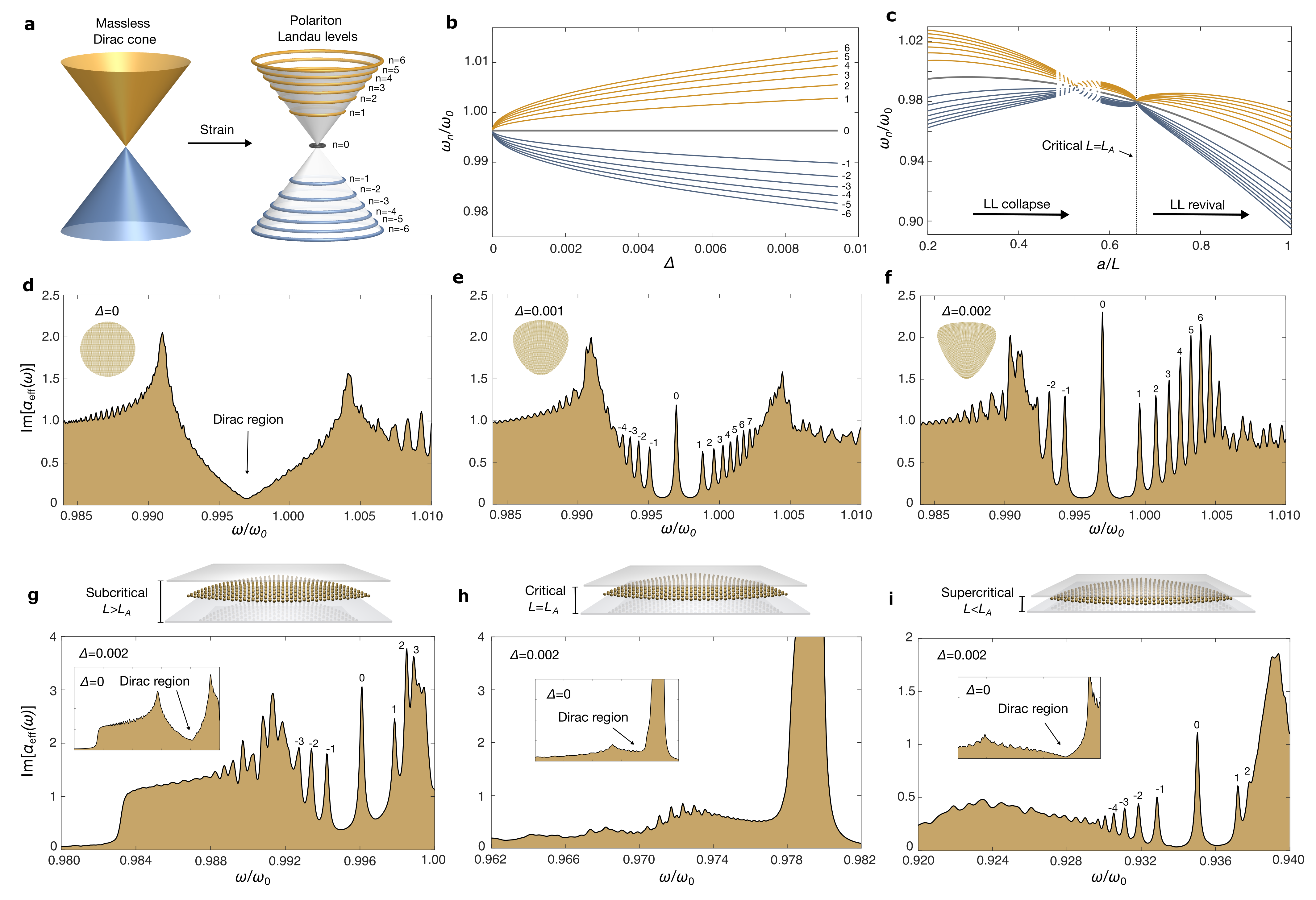}
\caption{\textbf{Cavity-induced collapse and revival of polariton Landau levels.} \textbf{a}  Schematic illustration of a massless Dirac cone splitting into quantized polariton Landau levels due to a large pseudo-magnetic field induced by strain. \textbf{b} Shows the squareroot dependence of the polariton Landau levels on the strain magnitude and Landau index for a fixed cavity height ($a/L=0.2$). \textbf{c} Shows the collapse and revival of the polariton Landau levels as the cavity height is reduced for a fixed strain magnitude ($\Delta=0.05$), as predicted by the effective Hamiltonian\,\eqref{eq:EffectHamiltonianK}. The dotted lines indicate the region of cavity heights where the linear Dirac cone approximation breaks down (see Supplementary Information for more details). \textbf{d}-\textbf{f} Local spectral function at the centre of the metasurface (see Methods for details) for a fixed cavity height ($a/L=0.2$), but different strain magnitudes of $\Delta=0$, $\Delta=0.001$, and $\Delta=0.002$, respectively. For the strained metasurface (see \textbf{e}), we observe a series of resonant peaks within the Dirac region which correspond to the polariton Landau levels (labeled according to their Landau index) which are not present in the unstrained case (compare with \textbf{d}). As the strain is increased (see \textbf{f}), the Landau level spacing increases in accordance with the analytical prediction (see \textbf{b}). \textbf{g}-\textbf{i} Local spectral function at the centre of the metasurface for a fixed strain magnitude ($\Delta=0.002$), but different cavity heights of $a/L=0.2$, $a/L=0.68$, and $a/L=1$, respectively. For subcritical cavity heights $L>L_{A}$ (see \textbf{g}), we observe Landau level peaks which are not present in the unstrained case (see inset). At the critical cavity height $L=L_{A}$ (see \textbf{h}), the pseudo-magnetic field is switched off and therefore no Landau level peaks are observed within the Dirac region; in fact, there is very little difference when compared to the corresponding unstrained metasurface (see inset). For supercritical cavity heights $L<L_{A}$ (see \textbf{i}), we observe Landau level peaks within the Dirac region which are not present in the unstrained metasurface (see inset), thus verifying the collapse and revival of the polariton Landau levels. Plots obtained with parameters $\Omega=0.01\omega_0$, $\Gamma=0.025\Omega$, and $\omega_0=c|\mathbf{K}|/2.2$. For \textbf{b}, \textbf{d}-\textbf{f} we use $\omega_0=c|\mathbf{K}|/3.5$ to maximize the size of the Dirac region. }
\label{fig:LDOSstrain}
\end{figure*}
%%TC:endignore

In what follows, we investigate some of the implications of the tunable pseudo-vector potential. Specifically, we consider a strain configuration described by the displacement field $\mathbf{u}(\mathbf{r})=(\Delta/a)(2xy\,,\,x^2-y^2)$ \cite{Guinea2009}, where $\Delta$ is a measure of the strain magnitude. This trigonal strain configuration gives rise to a vanishing pseudo-scalar potential and a pseudo-vector potential $\mathbf{A}=4(\Delta/a) A_{0} (y\,,\,-x)$, leading to a uniform pseudo-magnetic field $\mathbf{B}_{\tau }=\tau\curl\mathbf{A}=-\tau 8(\Delta/a) A_{0}\hat{\mathbf{z}}$ which, by virtue of time-reversal symmetry, has opposite signs for the $\mathrm{K}$ ($\tau=+1$) and $\mathrm{K}'$ ($\tau=-1$) valleys. 

Therefore, in the `semiclassical' limit \cite{Ashcroft1976}, polariton wavepackets propagating through the strained metasurface behave as if they were subjected to a Lorentz-like force $\mathbf{F}_{\tau}(L)=\operatorname{sign}(v_{\text{D}})\hat{\mathbf{v}}\cross \mathbf{B}_{\tau}$ which acts perpendicular to the group velocity direction $\hat{\mathbf{v}}$. Consequently, the polaritons exhibit cyclotron motion, as schematically depicted in figure\,\ref{fig:tunableorbits}a, in direct analogy with charged particles in real magnetic fields. Crucially, by modifying the nature of the dipole-dipole interactions via the cavity, one can tune the magnitude of the Lorentz-like force and also the effective cyclotron mass of the Dirac polaritons $m_c(L)=\delta\omega/v_{\text{D}}^2$, where $\delta\omega=\omega-\omega_{\text{D}}$ is the frequency relative to the Dirac point. As a result, in figure\,\ref{fig:tunableorbits}b we show how the corresponding cyclotron orbit radius $R_{\mathrm{c}}(L)=|m_c| v_{\text{D}}^2/|\mathbf{F}_{\tau}|$ can be tuned by varying only the cavity height. The dotted line indicates the region of cavity heights where the linear approximation of the Dirac cone breaks down (see Supplementary Information for more details). 

To verify the tunability of the cyclotron orbits, we simulated the evolution of Gaussian wavepackets using the effective Hamiltonian\,\eqref{eq:EffectHamiltonianK}, and in figure\,\ref{fig:tunableorbits}c we plot the trajectories of the centre-of-mass for a fixed strain magnitude, but different cavity heights (see Methods for details). As expected, the polariton wavepackets in the $\mathrm{K}$/$\mathrm{K}'$ valleys undergo cyclotron motion in opposite directions due to time-reversal symmetry (e.g., see snapshots along trajectory 5), and the orbit radii agree very well with the analytical predictions (see crosses in figure\,\ref{fig:tunableorbits}b). Note, the direction of the orbits depend on the signs of the Dirac velocity, cyclotron mass, and pseudo-magnetic field (see Supplementary Information for more details). For subcritical cavity heights $L>L_{A}$ where the short-range Coulomb interactions are dominant, the cyclotron radius expands as the cavity height is reduced (see trajectories 1-3). At the critical cavity height $L=L_{A}$, the pseudo-magnetic field is switched off and, as a result, the polariton wavepackets feel no Lorentz-like force despite the metasurface being strained (see trajectory 4). For supercritical cavity heights $L<L_{A}$ where the long-range photon-mediated interactions become dominant, the cyclotron orbits reemerge and the cyclotron radius now shrinks as the cavity height is reduced further (see trajectories 5-6). 

As the pseudo-magnetic field is increased one reaches the `quantum' limit, where the polariton cyclotron orbits undergo Landau quantization in direct analogy with charged particles in real magnetic fields \cite{Landau1930}. Therefore, as schematically depicted in figure\,\ref{fig:LDOSstrain}a, the massless Dirac cone collapses into a quantized Landau level spectrum $\omega_{n}(\Delta,L)=\omega_{\text{D}}^{}(L)\pm  \omega_{\mathrm{c}}(\Delta,L)\sqrt{n}$, where $n=0,1,2\dots$ is the Landau level index, and $\omega_{\mathrm{c}}(\Delta,L)=\sqrt{2|v_{\text{D}}||\mathbf{B}_{\tau}|}$ is the effective cyclotron frequency. In figure\,\ref{fig:LDOSstrain}b we show the characteristic squareroot dependence on the strain magnitude and Landau index \cite{Levy2010}, which is a manifestation of the pseudo-relativistic nature of the massless Dirac polaritons. In previous works, if one wanted to modify the Landau level spectrum in artificial lattices one had to fabricate an entirely new structure with a different strain configuration \cite{Rechtsman2013,Wen2019}. In stark contrast, here one can qualitatively tune the polariton Landau level spectrum by varying only the cavity height as shown in figure\,\ref{fig:LDOSstrain}c, where again the dotted line indicates the region of cavity heights where the linear Dirac cone approximation breaks down (see Supplementary Information for more details). Remarkably, despite the strain configuration being fixed, we predict an unprecedented collapse and revival of the polariton Landau levels by progressively decreasing the cavity height. 

To verify these analytical predictions, we go beyond the approximations of the effective Hamiltonian and calculate the effective polarizability $\alpha_{\text{eff}}(\omega)$ of a scatterer which fully accounts for the strong multiple scattering between the dipoles (see Methods for details). From this, we define a local spectral function $\operatorname{Im}[\alpha_{\text{eff}}(\omega)]$ which is related to the local density of states and characterizes the full spectral response of the metasurface. Figures\,\ref{fig:LDOSstrain}d-f show the local spectral function at the centre of the metasurface for a fixed cavity height ($L>L_{A}$), but different strain magnitudes (see Methods for details). For the strained metasurface (see figure\,\ref{fig:LDOSstrain}e), a series of resonant peaks emerge within the Dirac region which are not present in the unstrained case (see figure\,\ref{fig:LDOSstrain}d), and these directly correspond to the polariton Landau levels. As the strain is increased further (see figure\,\ref{fig:LDOSstrain}f), the spacing between the Landau level peaks increases in accordance with the analytical prediction. Interestingly, the Landau level peaks are much sharper than the resonance peak for a single scatterer, and their width is limited only by non-radiative losses. This is because the localized Landau level states are composed of large Fourier components that lie outside the light-line, which suppresses any resonant coupling to free-space photons, rendering the polariton Landau level states optically dark (see Supplementary Information for more details).

To verify the collapse and revival of the polariton Landau levels, in figures\,\ref{fig:LDOSstrain}g-i we show the local spectral function at the centre of the metasurface for a fixed strain configuration, but different cavity heights (see Methods for details). For subcritical cavity heights $L>L_{A}$ where the short-range Coulomb interactions are dominant (see figure\,\ref{fig:LDOSstrain}g), we clearly observe the presence of Landau level peaks within the Dirac region which are not present in the unstrained case (see inset). At the critical cavity height $L=L_{A}$ (see figure\,\ref{fig:LDOSstrain}h), the pseudo-magnetic field is switched off and, as a result, no Landau level peaks can be observed within the Dirac region despite the applied strain (compare with inset showing the strain-free case), thus verifying the collapse of the polariton Landau levels. For supercritical cavity heights $L<L_{A}$ where the long-range photon-mediated interactions become dominant (see figure\,\ref{fig:LDOSstrain}i), we indeed observe the revival of the polariton Landau level peaks within the Dirac region which are not present in the unstrained case (see inset).

This work poses new intriguing questions beyond the realm of conventional tight-binding models, which can be explored in a variety of experimental platforms across the electromagnetic spectrum. For example, do quantum-Hall-like edge states associated with the polariton Landau levels exist in the presence of long-range interactions? Can one controllably switch them on and off, or tune their properties, by modifying the nature of the dipole-dipole interactions? Moreover, do novel edge states emerge at interfaces separating regions with different photonic environments? The polariton Landau levels also provide a novel way of sculpting the local density of states --- this could be exploited for enhancing/suppressing spontaneous emission and strong coupling effects for emitters in subwavelength photonic structures. Furthermore, while we have demonstrated a tunable Lorentz-like force and tunable Landau levels, one should also be able to observe Aharonov-Bohm-like interference patterns which will depend qualitatively on the surrounding photonic environment. Finally, here we have focused on some of the implications of a tunable pseudo-magnetic field, but what are the implications of a tunable pseudo-electric field arising from the pseudo-scalar potential? To conclude, while intense efforts are devoted towards designing systems that emulate tight-binding models, this work hints towards a richer landscape of physics emerging from long-range interactions which are prevalent in electromagnetic systems.

%%TC:ignore

%=============================================
\begin{scriptsize}
\section{methods}
%=============================================

\noindent The unstrained metasurface is composed of a honeycomb array of resonant point-dipole scatterers located at periodic positions $\mathbf{R}_\mathrm{A}=\mathbf{R}-\mathbf{d}/2$ and $\mathbf{R}_\mathrm{B}=\mathbf{R}+\mathbf{d}/2$ which form the $\mathrm{A}$ and $\mathrm{B}$ hexagonal sublattices, respectively. Here, $\mathbf{d}=a(0,1)$ is the vector connecting the $\mathrm{A}$ and $\mathrm{B}$ sites in the unit cell, and $\mathbf{R}=l_1\mathbf{a}_1+l_2\mathbf{a}_2$ are the set of lattice vectors, where $l_1$ and $l_2$ are integers, and $\mathbf{a}_1^{}=(\sqrt{3}a/2)(-1,\sqrt{3})$ and $\mathbf{a}_2^{}=(\sqrt{3}a/2)(1,\sqrt{3})$ are the primitive lattice vectors. The corresponding reciprocal lattice vectors are $\mathbf{G}=n_1 \mathbf{b}_1 +n_2 \mathbf{b}_2$, where $n_1$ and $n_2$ are integers, and $\mathbf{b}_1=(2\pi/3a)(-\sqrt{3},1)$ and $\mathbf{b}_2=(2\pi/3a)(\sqrt{3},1)$ are the primitive reciprocal lattice vectors. The nearest-neighbor vectors which connect dipoles located on different sublattices are given by $\mathbf{e}_1^{}=a(0,-1)$, $\mathbf{e}_2^{}=(a/2)(\sqrt{3},1)$, and $\mathbf{e}_3^{}=(a/2)(-\sqrt{3},1)$, and the next-nearest-neighbor vectors which connect dipoles located on the same sublattice are given by $\mathbf{a}_1^{}$, $\mathbf{a}_2^{}$, $\mathbf{a}_3^{}=\mathbf{a}_2^{}-\mathbf{a}_1^{}$, $\mathbf{a}_4^{}=-\mathbf{a}_1^{}$, $\mathbf{a}_5^{}=-\mathbf{a}_2^{}$, and $\mathbf{a}_6^{}=-\mathbf{a}_3^{}$. When strain is applied to the metasurface the dipoles are displaced to aperiodic positions given by $\bar{\mathbf{R}}_{\mathrm{A}/\mathrm{B}}=\mathbf{R}_{\mathrm{A}/\mathrm{B}}+\mathbf{u}(\mathbf{R}_{\mathrm{A}/\mathrm{B}})$. Finally, the metasurface is embedded at the centre of a planar photonic cavity formed by two perfectly reflecting mirrors located at $z=\pm L/2$.

%=============================================
\subsection{Effective polariton Hamiltonian }
%=============================================

\noindent  In this section we use a coupled-dipole model to derive an effective Hamiltonian describing the polaritons near the $\mathrm{K}$/$\mathrm{K}'$ points, which is valid to leading order in strain. The scatterers are assumed to be anisotropic such that the induced dipole moments point normal to the plane of the metasurface, i.e., their bare polarizability tensor reads $\bm{\alpha}_0(\omega)=\alpha_0(\omega)\hat{\mathbf{z}}\hat{\mathbf{z}}$, where $\alpha_0(\omega)$ is defined in the main text. To describe a scatterer's response to an external field, one has to modify this bare polarizability to take into account the interaction with its own scattered field. The renormalized polarizability of a scatterer inside the cavity therefore reads
%---------------------
\begin{equation}
    \alpha_{}^{-1}(\omega)=\alpha_{0}^{-1}(\omega)-\Sigma(\omega)\,,
    \label{eq:RadiativeCorrection1}
\end{equation}
%---------------------
where the polarizability correction is given by
%---------------------
\begin{equation}
\begin{split}
    \Sigma(\omega)=\mathrm{i}\frac{2a^3k_{\omega}^3}{3}+\frac{4a^3}{L^3}\left[\operatorname{Li}_{3}\left(\mathrm{e}^{\mathrm{i}k_{\omega}L}\right)-\mathrm{i}k_{\omega}L\operatorname{Li}_{2}\left(\mathrm{e}^{\mathrm{i}k_{\omega} L}\right)\right]\,,
    \label{eq:RadiativeCorrection}
\end{split}
\end{equation}
%-------------------------------------
where $k_{\omega}=\omega/c$ and $\operatorname{Li}_{n}(z)=\sum_{l=1}^{\infty}z^l/l^n$ is the polylogarithm of order $n$ (see Supplementary Information for the derivation). The first term in equation\,\eqref{eq:RadiativeCorrection} is the usual radiative-reaction correction in free-space \cite{Wokaun1982}, whereas the other terms encode the corrections due to the cavity. In the absence of a driving field, the induced dipole moments $\mathbf{p}_{\mathbf{R}_{\mathrm{A}}}^{}=p_{\mathbf{R}_{\mathrm{A}}}^{}\hat{\mathbf{z}}$ and $\mathbf{p}_{\mathbf{R}_{\mathrm{B}}}^{}=p_{\mathbf{R}_{\mathrm{B}}}^{}\hat{\mathbf{z}}$ on the $\mathrm{A}$ and $\mathrm{B}$ sublattices are given by the self-consistent coupled-dipole equations
%---------------------
\begin{equation}
  \frac{1}{\alpha_{}(\omega)}p_{\mathbf{R}_{\mathrm{A}}}^{}(\omega)=\sum_{\mathclap{\mathbf{R}'_{\mathrm{A}}\neq\mathbf{R}_{\mathrm{A}}^{}}}\mathsf{G}^{}_{}(\bar{\mathbf{R}}_{\mathrm{A}}^{}-\bar{\mathbf{R}}'_{\mathrm{A}},\omega)p_{\mathbf{R}'_{\mathrm{A}}}(\omega)+\sum_{\mathbf{R}_{\mathrm{B}}^{}}\mathsf{G}^{}_{}(\bar{\mathbf{R}}_{\mathrm{A}}^{}-\bar{\mathbf{R}}_{\mathrm{B}}^{},\omega)p_{\mathbf{R}_{\mathrm{B}}}^{}(\omega)
  \label{eq:CDM1}
\end{equation}
%---------------------
and
%---------------------
\begin{equation}
  \frac{1}{\alpha_{}(\omega)}p_{\mathbf{R}_{\mathrm{B}}}^{}(\omega)=\sum_{\mathclap{\mathbf{R}'_{\mathrm{B}}\neq\mathbf{R}_{\mathrm{B}}^{}}}\mathsf{G}^{}_{}(\bar{\mathbf{R}}_{\mathrm{B}}^{}-\bar{\mathbf{R}}'_{\mathrm{B}},\omega)p_{\mathbf{R}'_{\mathrm{B}}}(\omega)+\sum_{\mathbf{R}_{\mathrm{A}}^{}}\mathsf{G}^{}_{}(\bar{\mathbf{R}}_{\mathrm{B}}^{}-\bar{\mathbf{R}}_{\mathrm{A}}^{},\omega)p_{\mathbf{R}_{\mathrm{A}}}^{}(\omega)\,,
  \label{eq:CDM2}
\end{equation}
%---------------------
where the first and second sums in each equation correspond to the field ($z$-component) generated by all the other dipoles belonging to the same and opposite sublattice, respectively (see Supplementary Information for the derivation). The interactions between the dipoles are mediated by the cavity Green's function which can be expressed in the spectral domain as
%-------------------------------------
\begin{equation}
\begin{split}
    \mathsf{G}(\mathbf{r},\omega )=\mathrm{i}\frac{\pi a ^3 k_{\omega}^2}{L} \sum_{m=-\infty}^{\infty}\left(1-\frac{k_{m}^2}{k_{\omega}^2} \right) \operatorname{H}_{0}^{(1)}\left(r \sqrt{k_{\omega}^2-k_{m}^2}\right),
\end{split}
\label{eq:FullCavityGreensFunction}
\end{equation}
%-------------------------------------
where $k_{m}=2 m\pi/L$ and $\operatorname{H}_0^{(1)}$ is the Hankel function of zeroth order and first kind (see Supplementary Information for the derivation). Note, equation\,\eqref{eq:FullCavityGreensFunction} is the $zz$-component of the dyadic Green's function where both the source and observation points are located at the centre of the cavity and connected by the in-plane vector $\mathbf{r}$. 

We decompose the cavity Green's function\,\eqref{eq:FullCavityGreensFunction} into its longitudinal and transverse components $\mathsf{G}=\mathsf{G}^{}_{\parallel}+\mathsf{G}^{}_{\perp}$, where the longitudinal component can be expressed in the spectral domain as 
%---------------------
\begin{equation}
    \mathsf{G}^{}_{\parallel}(\mathbf{r})=-\frac{4a^3}{L}\sum_{m=1}^{\infty}k_{m}^2\operatorname{K}_0(k_{m}r)\,,
    \label{eq:CoulombGreensFunction}
\end{equation}
%---------------------
where $\operatorname{K}_0^{}$ is the modified Bessel function of zeroth order and second kind (see Supplementary Information for the derivation). The longitudinal component of the Green's function\,\eqref{eq:CoulombGreensFunction} mediates Coulomb interactions between the dipoles inside the cavity. In contrast, the transverse component  $\mathsf{G}_{\perp}=\mathsf{G}-\mathsf{G}_{\parallel}$ describes interactions that are mediated by the transverse photonic modes of the cavity. For simplicity, we retain only the dominant contribution from the fundamental transverse electromagnetic (TEM) mode of the cavity which has a linear dispersion $\omega_{\mathbf{k}}=c|\mathbf{k}|$ and polarization (along $\hat{\mathbf{z}}$) that are independent of the cavity height. This is a good approximation since we are interested in the regime of small cavity heights where all the other cavity modes are detuned from the dipole resonances. The corresponding Green's function for the TEM mode reads
%---------------------------------------------
\begin{equation}
    {\mathsf{G}}_{\perp}^{\text{TEM}}(\mathbf{r},\omega)=\frac{ 2 a^3 k_{\omega}^2}{ L}\int\frac{\mathrm{d}^2\mathbf{k}}{2\pi}\frac{\mathrm{e}^{\mathrm{i}\mathbf{k}\cdot\mathbf{r}}}{k^2-k_{\omega}^2-\mathrm{i}0^{+}}=\frac{\mathrm{i}\pi a^3 k_{\omega}^2}{ L}\operatorname{H}_0^{(1)}\left(k_{\omega}r\right)\,,
    \label{eq:TEMGreensFunction}
\end{equation}
%---------------------------------------------
where the infinitesimal imaginary shift $\mathrm{i}0^{+}$ ensures that the dipoles emit outgoing waves (see Supplementary Information for the derivation).

In what follows, we derive the effective Hamiltonian for the $\mathrm{K}$ valley since the effective Hamiltonian for the $\mathrm{K}'$ valley is related via time-reversal symmetry. To analyze the polaritons near the $\mathrm{K}$ point, we write the dipole moments as 
%---------------------
\begin{equation}
    p_{\mathbf{R}_\mathrm{A}}^{}(\omega)=\mathrm{e}^{\mathrm{i}\mathbf{K}\cdot\mathbf{R}_\mathrm{A}}\psi^{\mathrm{K}}_{\mathrm{A}}(\mathbf{R}_\mathrm{A},\omega)\quad,\quad p_{\mathbf{R}_\mathrm{B}}^{}(\omega)=\mathrm{e}^{\mathrm{i}\mathbf{K}\cdot\mathbf{R}_\mathrm{B}}\psi^{\mathrm{K}}_{\mathrm{B}}(\mathbf{R}_\mathrm{B},\omega)\,,
    \label{eq:EnvelopeFunctions}
\end{equation}
%---------------------
where $\psi^{\mathrm{K}}_{\mathrm{A}}$ and $\psi^{\mathrm{K}}_{\mathrm{B}}$ are slowly-varying envelope fields for the $\mathrm{A}$ and $\mathrm{B}$ sublattices, respectively. Here we have used the crystal reference frame \cite{deJuan2013} where the dipoles are located at their original unstrained positions. To not overburden notation, we suppress the valley index until the end. By introducing the Fourier transform of the envelope fields
%---------------------------------------------
\begin{equation}
\widetilde{\psi}^{}_{\mathrm{A}}(\mathbf{k},\omega)=\int\;\frac{\mathrm{d}^2\mathbf{r}}{2\pi}\psi^{}_{\mathrm{A}}(\mathbf{r},\omega)\mathrm{e}^{-\mathrm{i}\mathbf{k}\cdot\mathbf{r}}\,,\,\widetilde{\psi}^{}_{\mathrm{B}}(\mathbf{k},\omega)=\int\;\frac{\mathrm{d}^2\mathbf{r}}{2\pi}\psi^{}_{\mathrm{B}}(\mathbf{r},\omega)\mathrm{e}^{-\mathrm{i}\mathbf{k}\cdot\mathbf{r}}
\label{eq:FTEnvelopeFunctions}
\end{equation}
%---------------------------------------------
we can write the coupled-dipole equations\,\eqref{eq:CDM1} and \eqref{eq:CDM2} in matrix form as
%---------------------------------------------
\begin{equation}
\frac{1}{\check{\alpha}(\omega)}\widetilde{\psi}(\mathbf{k},\omega)=\;\int\mathrm{d}^2\mathbf{k}'\,\big[\mathcal{D}_{\parallel}^{}(\mathbf{k},\mathbf{k}')+\mathcal{D}_{\perp}^{}(\mathbf{k},\mathbf{k}',\omega)\big]\widetilde{\psi}(\mathbf{k}',\omega)\,,
\label{eq:CDMMatrixForm}
\end{equation}
%---------------------------------------------
where $\check{\alpha}_{}^{-1}(\omega)=\alpha_{0}^{-1}(\omega)-\operatorname{Re}[\Sigma(\omega)]$. In equation\,\eqref{eq:CDMMatrixForm}, the Fourier transform of the spinor envelope field reads $\widetilde{\psi}(\mathbf{k},\omega)=[\widetilde{\psi}_{\mathrm{A}}(\mathbf{k},\omega)\,,\,\widetilde{\psi}_{\mathrm{B}}(\mathbf{k},\omega)]^{\mathrm{T}}$, the longitudinal dynamical matrix encoding the Coulomb interactions is
%---------------------------------------------
\begin{equation}
\mathcal{D}_{\parallel}^{}(\mathbf{k},\mathbf{k}')=\begin{pmatrix}\mathcal{D}_{\parallel}^{\text{AA}}(\mathbf{k},\mathbf{k}') & \mathcal{D}_{\parallel}^{\text{AB}}(\mathbf{k},\mathbf{k}')\\
\mathcal{D}_{\parallel}^{\text{AB} *}(\mathbf{k}',\mathbf{k})  & \mathcal{D}_{\parallel}^{\text{BB}}(\mathbf{k},\mathbf{k}')\end{pmatrix}\,, 
\label{eq:LongitudinalDynamicMatrix}
\end{equation}
%---------------------------------------------
and the transverse dynamical matrix encoding the photon-mediated interactions reads
%---------------------
\begin{equation}
\mathcal{D}_{\perp}^{}(\mathbf{k},\mathbf{k}',\omega)=\begin{pmatrix}\mathcal{D}_{\perp}^{\text{AA}}(\mathbf{k},\mathbf{k}',\omega) & \mathcal{D}_{\perp}^{\text{AB}}(\mathbf{k},\mathbf{k}',\omega)\\
\mathcal{D}_{\perp}^{\text{AB} *}(\mathbf{k}',\mathbf{k},\omega)  & \mathcal{D}_{\perp}^{\text{BB}}(\mathbf{k},\mathbf{k}',\omega)\end{pmatrix}\,.
\label{eq:TransverseDynamicMatrix}
\end{equation}
%---------------------

We will first analyze the longitudinal dynamical matrix elements in equation\,\eqref{eq:LongitudinalDynamicMatrix}, where the intersublattice (off-diagonal) matrix elements read
%---------------------
\begin{equation}
\begin{split}
    \mathcal{D}_{\parallel}^{\text{AB}}(\mathbf{k},\mathbf{k}')=\int\frac{\mathrm{d}^2\mathbf{r}}{(2\pi)^2}\,\sum_{\mathbf{R}}&\mathsf{G}^{}_{\parallel}(\mathbf{R}-\mathbf{d}+\mathbf{u}(\mathbf{r})-\mathbf{u}(\mathbf{r}-\mathbf{R}+\mathbf{d}))\\
    &\times\mathrm{e}^{-\mathrm{i}(\mathbf{K}+\mathbf{k}')\cdot (\mathbf{R}-\mathbf{d})}\mathrm{e}^{\mathrm{i}(\mathbf{k}'-\mathbf{k})\cdot \mathbf{r}}\,.
  \end{split}
  \label{eq:DABLongitudinal1}
\end{equation}
%---------------------
Since the Coulomb interactions are short-range, the lattice sums converge rapidly and therefore we retain only the dominant nearest and next-nearest-neighbor Coulomb interactions. Then, after expanding equation\,\eqref{eq:DABLongitudinal1} to leading order in the displacement field we obtain
%---------------------
\begin{equation}
\begin{split}
    \mathcal{D}_{\parallel}^{\text{AB}}&(\mathbf{k},\mathbf{k}')=\int\frac{\mathrm{d}^2\mathbf{r}}{(2\pi)^2}\,\sum_{m=1}^{3}\mathsf{G}^{}_{\parallel}(\mathbf{e}_m)\mathrm{e}^{-\mathrm{i}(\mathbf{K}+\mathbf{k}')\cdot \mathbf{e}_m}\mathrm{e}^{\mathrm{i}(\mathbf{k}'-\mathbf{k})\cdot \mathbf{r}}\\
    &\times\Bigg[1+\frac{\beta_{\text{nn}}}{2\pi}\frac{e_m^i}{|\mathbf{e}_m|^2}\int\mathrm{d}^2\mathbf{q}\,\widetilde{u}_i(\mathbf{q})\big(1-\mathrm{e}^{-\mathrm{i}\mathbf{q}\cdot \mathbf{e}_m}\big)\mathrm{e}^{\mathrm{i}\mathbf{q}\cdot \mathbf{r}}\Bigg]\,,
  \end{split}
  \label{eq:DABLongitudinal2}
\end{equation}
%---------------------
where $\beta_{\text{nn}}=|\partial \log [\mathsf{G}_{\parallel}(\mathbf{e}_m)]/\partial \log (|\mathbf{e}_m|)|$ encodes how the Coulomb interaction strength changes with respect to changes in the nearest-neighbor separation distance, and
%---------------------------------------------
\begin{equation}
 \widetilde{\mathbf{u}}(\mathbf{q})=\int_{}\frac{\mathrm{d}^2\mathbf{r}}{2\pi}\, \mathbf{u}(\mathbf{r}) \mathrm{e}^{-\mathrm{i} \mathbf{q}\cdot\mathbf{r}}\,
 \label{eq:FTDisplacementField}
\end{equation}
%---------------------------------------------
is the Fourier transform of the displacement field. Next, performing the spatial integral in equation\,\eqref{eq:DABLongitudinal2} we obtain
%---------------------
\begin{equation}
\begin{split}
    \mathcal{D}_{\parallel}^{\text{AB}}&(\mathbf{k},\mathbf{k}')=\sum_{m=1}^{3}\mathsf{G}^{}_{\parallel}(\mathbf{e}_m)\mathrm{e}^{-\mathrm{i}(\mathbf{K}+\mathbf{k}')\cdot \mathbf{e}_m}\\
    &\times\Bigg\{\delta(\mathbf{k}'-\mathbf{k})+\frac{\beta_{\text{nn}}}{2\pi}\frac{e_m^i}{|\mathbf{e}_m|^2} \widetilde{u}_i(\mathbf{k}-\mathbf{k}')\bigg[1-\mathrm{e}^{-\mathrm{i}(\mathbf{k}-\mathbf{k}')\cdot \mathbf{e}_m}\bigg]\Bigg\}\,.
  \end{split}
  \label{eq:DABLongitudinal3}
\end{equation}
%---------------------
Finally, expanding equation\,\eqref{eq:DABLongitudinal3} to leading order in $\mathbf{k}$ and $\mathbf{k}'$ we obtain
%---------------------
\begin{equation}
\begin{split}
    \mathcal{D}_{\parallel}^{\text{AB}}&(\mathbf{k},\mathbf{k}')=\sum_{m=1}^{3}\mathsf{G}^{}_{\parallel}(\mathbf{e}_m)\mathrm{e}^{-\mathrm{i}\mathbf{K}\cdot \mathbf{e}_m}\\
    &\times\Bigg[\delta(\mathbf{k}'-\mathbf{k})(1-\mathrm{i}e_m^ik'_i)+\frac{\beta_{\text{nn}}}{2\pi}\frac{e_m^i e_m^j}{|\mathbf{e}_m|^2}  \widetilde{\nabla u}_{ij}(\mathbf{k}-\mathbf{k}')\Bigg]\,,
  \end{split}
  \label{eq:DABLongitudinal4}
\end{equation}
%---------------------
where we have identified $\mathrm{i}k_i\widetilde{u}_j(\mathbf{k})=\widetilde{\nabla u}_{ij}(\mathbf{k})$, which is the Fourier transform of the displacement gradient tensor defined as $\nabla u_{ij}=\partial u_j/\partial r_i$. Performing similar analysis for the intrasublattice (diagonal) matrix elements in equation\,\eqref{eq:LongitudinalDynamicMatrix} we obtain
%---------------------
\begin{equation}
\begin{split}
    &\mathcal{D}_{\parallel}^{\text{AA}/\text{BB}}(\mathbf{k},\mathbf{k}')=\sum_{m=1}^{6}\mathsf{G}^{}_{\parallel}(\mathbf{a}_m)\mathrm{e}^{-\mathrm{i}\mathbf{K}\cdot \mathbf{a}_m}\\
    &\times\Bigg[\delta(\mathbf{k}'-\mathbf{k})(1-\mathrm{i}a_m^ik'_i)+\frac{\beta_{\text{nnn}}}{2\pi}\frac{a_m^i a_m^j}{|\mathbf{a}_m|^2}  \widetilde{\nabla u}_{ij}(\mathbf{k}-\mathbf{k}')\Bigg]\,,
  \end{split}
  \label{eq:DAALongitudinal1}
\end{equation}
%---------------------
where $\beta_{\text{nnn}}=|\partial \log [\mathsf{G}_{\parallel}(\mathbf{a}_m)]/\partial \log (|\mathbf{a}_m|)|$ encodes how the Coulomb interaction strength changes with respect to changes in the next-nearest-neighbor separation distance. 

We will now analyze the transverse dynamical matrix elements in equation\,\eqref{eq:TransverseDynamicMatrix}, where the intersublattice (off-diagonal) matrix elements read
%---------------------
\begin{equation}
\begin{split}
    \mathcal{D}_{\perp}^{\text{AB}}(\mathbf{k},\mathbf{k}',\omega)=\int\frac{\mathrm{d}^2\mathbf{r}}{(2\pi)^2}\,\sum_{\mathbf{R}}&\mathsf{G}^{\text{TEM}}_{\perp}(\mathbf{R}-\mathbf{d}+\mathbf{u}(\mathbf{r})-\mathbf{u}(\mathbf{r}-\mathbf{R}+\mathbf{d}),\omega)\\
    &\times\mathrm{e}^{-\mathrm{i}(\mathbf{K}+\mathbf{k}')\cdot (\mathbf{R}-\mathbf{d})}\mathrm{e}^{\mathrm{i}(\mathbf{k}'-\mathbf{k})\cdot \mathbf{r}}\,.
  \end{split}
  \label{eq:DABTransverse1}
\end{equation}
%---------------------
Since the photon-mediated interactions are long-range, we seek to perform the lattice sums in reciprocal space where they converge rapidly. To do this, we first insert the integral representation of the TEM Green's function\,\eqref{eq:TEMGreensFunction} into equation\,\eqref{eq:DABTransverse1} which gives
%---------------------
\begin{equation}
\begin{split}
    \mathcal{D}_{\perp}^{\text{AB}}(\mathbf{k},\mathbf{k}',\omega)&=\int\frac{\mathrm{d}^2\mathbf{r}}{(2\pi)^2}\int\frac{\mathrm{d}^2\mathbf{k}''}{2\pi}\,\sum_{\mathbf{R}}\,\widetilde{\mathsf{G}}^{\text{TEM}}_{\perp}(\mathbf{k}'',\omega)\mathrm{e}^{\mathrm{i}(\mathbf{k}'-\mathbf{k})\cdot \mathbf{r}}\\
    &\times\mathrm{e}^{\mathrm{i}(\mathbf{k}''-\mathbf{k}'-\mathbf{K})\cdot(\mathbf{R}-\mathbf{d})}\mathrm{e}^{\mathrm{i}\mathbf{k}''\cdot[\mathbf{u}(\mathbf{r})-\mathbf{u}(\mathbf{r}-\mathbf{R}+\mathbf{d})]}\,,
  \end{split}
  \label{eq:DABTransverse2}
\end{equation}
%---------------------
where $\widetilde{\mathsf{G}}^{\text{TEM}}_{\perp}(\mathbf{k},\omega)=(2a^3 k_{\omega}^2/ L)(k^2-k_{\omega}^2)^{-1}$
is the Fourier transform of the TEM Green's function. Next, the expansion of equation\,\eqref{eq:DABTransverse2} 
to leading order in the displacement field then yields
%---------------------
\begin{equation}
\begin{split}
    &\mathcal{D}_{\perp}^{\text{AB}}(\mathbf{k},\mathbf{k}',\omega)=\int\frac{\mathrm{d}^2\mathbf{r}}{(2\pi)^2}\int\frac{\mathrm{d}^2\mathbf{k}''}{2\pi}\,\sum_{\mathbf{R}}\,\widetilde{\mathsf{G}}^{\text{TEM}}_{\perp}(\mathbf{k}'',\omega)\mathrm{e}^{\mathrm{i}(\mathbf{k}'-\mathbf{k})\cdot \mathbf{r}}\\
    &\times\mathrm{e}^{\mathrm{i}(\mathbf{k}''-\mathbf{k}'-\mathbf{K})\cdot(\mathbf{R}-\mathbf{d})}\Bigg\{1+\frac{\mathrm{i}}{2\pi}k''_i\int\mathrm{d}^2\mathbf{q}\,\widetilde{u}_i(\mathbf{q})\bigg[1-\mathrm{e}^{-\mathrm{i}\mathbf{q}\cdot (\mathbf{R}-\mathbf{d})}\bigg]\mathrm{e}^{\mathrm{i}\mathbf{q}\cdot \mathbf{r}}\Bigg\}\,,
  \end{split}
  \label{eq:DABTransverse3}
\end{equation}
%---------------------
and after performing the spatial integral in equation\,\eqref{eq:DABTransverse3} we obtain
%---------------------
\begin{equation}
\begin{split}
    &\mathcal{D}_{\perp}^{\text{AB}}(\mathbf{k},\mathbf{k}',\omega)=\int\frac{\mathrm{d}^2\mathbf{k}''}{2\pi}\,\sum_{\mathbf{R}}\,\widetilde{\mathsf{G}}^{\text{TEM}}_{\perp}(\mathbf{k}'',\omega)\mathrm{e}^{\mathrm{i}(\mathbf{k}''-\mathbf{k}'-\mathbf{K})\cdot(\mathbf{R}-\mathbf{d})}\\
    &\times\Bigg\{\delta(\mathbf{k}'-\mathbf{k})+\frac{\mathrm{i}}{2\pi}k''_i\,\widetilde{u}_i(\mathbf{k}-\mathbf{k}')\bigg[1-\mathrm{e}^{-\mathrm{i}(\mathbf{k}-\mathbf{k}')\cdot (\mathbf{R}-\mathbf{d})}\bigg]\Bigg\}\,.
  \end{split}
  \label{eq:DABTransverse4}
\end{equation}
%---------------------
We now use Poisson's summation identity $\sum_{\mathbf{R}}\exp[\mathrm{i}(\mathbf{k}'-\mathbf{k})\cdot\mathbf{R}]=[(2\pi)^2/\mathcal{A}]\sum_{\mathbf{G}}\delta(\mathbf{k}'-\mathbf{k}+\mathbf{G})$ to convert the sum over lattice vectors to a sum over reciprocal lattice vectors which gives
%---------------------
\begin{equation}
\begin{split}
    &\mathcal{D}_{\perp}^{\text{AB}}(\mathbf{k},\mathbf{k}',\omega)=\frac{2\pi}{\mathcal{A}}\sum_{\mathbf{G}}\phi_{\mathbf{G}}^{}\Bigg\{\widetilde{\mathsf{G}}^{\text{TEM}}_{\perp}(\mathbf{k}'+\mathbf{K}-\mathbf{G},\omega)\delta(\mathbf{k}'-\mathbf{k})\\
    &+\frac{\mathrm{i}}{2\pi}\bigg[\widetilde{\mathsf{G}}^{\text{TEM}}_{\perp}(\mathbf{k}'+\mathbf{K}-\mathbf{G},\omega)(\mathbf{k}'+\mathbf{K}-\mathbf{G})_i-(\mathbf{k}'\leftrightarrow\mathbf{k})\bigg]\widetilde{u}_i(\mathbf{k}-\mathbf{k}')\Bigg\}\,,
  \end{split}
  \label{eq:DABTransverse5}
\end{equation}
%---------------------
where $\phi_{\mathbf{G}}^{}=\exp(\mathrm{i}\mathbf{G}\cdot\mathbf{d})$ are non-trivial phase factors that are crucial for maintaining the correct symmetry. Finally, expanding equation\,\eqref{eq:DABTransverse5} to leading order in $\mathbf{k}$ and $\mathbf{k}'$ we obtain
%---------------------
\begin{equation}
\begin{split}
    &\mathcal{D}_{\perp}^{\text{AB}}(\mathbf{k},\mathbf{k}',\omega)=\sum_{\mathbf{G}}\frac{\omega^2\xi^2 \phi_{\mathbf{G}}^{}}{\omega_{\mathbf{K}-\mathbf{G}}^2-\omega^2}\Bigg\{\delta(\mathbf{k}'-\mathbf{k})-\frac{1}{2\pi}\widetilde{\nabla u}_{ii}(\mathbf{k}'-\mathbf{k})\\
    &-\frac{2c^2 (\mathbf{K}-\mathbf{G})_i}{\omega_{\mathbf{K}-\mathbf{G}}^2-\omega^2}\bigg[k'_i \delta(\mathbf{k}'-\mathbf{k})-\frac{1}{2\pi}(\mathbf{K}-\mathbf{G})_j\widetilde{\nabla u}_{ij}(\mathbf{k}-\mathbf{k}')\bigg]\Bigg\}\,,
  \end{split}
  \label{eq:DABTransverse6}
\end{equation}
%---------------------
where $\xi^2=4\pi a^3/\mathcal{A}L$ parameterizes the strength of the light-matter interaction. Performing similar analysis for the intrasublattice (diagonal) matrix elements in equation\,\eqref{eq:TransverseDynamicMatrix} we obtain
%---------------------
\begin{equation}
\begin{split}
    &\mathcal{D}_{\perp}^{\text{AA}/\text{BB}}(\mathbf{k},\mathbf{k}',\omega)=\sum_{\mathbf{G}}\frac{\omega^2\xi^2 }{\omega_{\mathbf{K}-\mathbf{G}}^2-\omega^2}\Bigg\{\delta(\mathbf{k}'-\mathbf{k})-\frac{1}{2\pi}\widetilde{\nabla u}_{ii}(\mathbf{k}'-\mathbf{k})\\
    &-\frac{2c^2 (\mathbf{K}-\mathbf{G})_i}{\omega_{\mathbf{K}-\mathbf{G}}^2-\omega^2}\bigg[k'_i \delta(\mathbf{k}'-\mathbf{k})-\frac{1}{2\pi}(\mathbf{K}-\mathbf{G})_j\widetilde{\nabla u}_{ij}(\mathbf{k}-\mathbf{k}')\bigg]\Bigg\}\\
    &-\operatorname{Re}[\mathsf{G}_{\perp}^{\text{TEM}}(0,\omega)]\delta(\mathbf{k}'-\mathbf{k})\,.
  \end{split}
  \label{eq:DAATransverse1}
\end{equation}
%---------------------

To obtain an effective Hamiltonian for the polariton envelope fields, we first neglect non-radiative losses for simplicity and assume small detuning such that $\omega_0^2-\omega^2\simeq2\omega_0(\omega_0-\omega)$. Next, we neglect the frequency dependence of the polarizability correction\,\eqref{eq:RadiativeCorrection} and the transverse dynamical matrix\,\eqref{eq:TransverseDynamicMatrix} by evaluating them at the free-space resonant frequency $\omega_0$. This approximation is equivalent to treating the light-matter interactions to second order in perturbation theory \cite{Craig1984}, and is valid for polaritons near the corners of the Brillouin zone where $\omega_0\Omega\xi^2\ll \omega_{\mathbf{K}-\mathbf{G}}^2-\omega_0^2$. Therefore, by Fourier transforming equation\,\eqref{eq:CDMMatrixForm} to the real-space and time domains, we obtain the equation of motion $\mathrm{i}\partial_t\psi_{\mathrm{K}}(\mathbf{r},t)=\mathcal{H}_{\mathrm{K}}\psi_{\mathrm{K}}(\mathbf{r},t)$ for the spinor envelope field in the $\mathrm{K}$ valley  $\psi_{\mathrm{K}}(\mathbf{r},t)=[\psi^{\mathrm{K}}_{\mathrm{A}}(\mathbf{r},t)\,,\,\psi^{\mathrm{K}}_{\mathrm{B}}(\mathbf{r},t)]^{\mathrm{T}}$, where the effective Hamiltonian is given by equation\,\eqref{eq:EffectHamiltonianK} in the main text. Similarly, we obtain the equation of motion $\mathrm{i}\partial_t\psi_{\mathrm{K}'}(\mathbf{r},t)=\mathcal{H}_{\mathrm{K}'}\psi_{\mathrm{K}'}(\mathbf{r},t)$ for the spinor envelope field in the $\mathrm{K}'$ valley $\psi_{\mathrm{K}'}(\mathbf{r},t)=[\psi^{\mathrm{K}'}_{\mathrm{A}}(\mathbf{r},t)\,,\,\psi^{\mathrm{K}'}_{\mathrm{B}}(\mathbf{r},t)]^{\mathrm{T}}$, where the effective Hamiltonian is related to equation\,\eqref{eq:EffectHamiltonianK} by time-reversal symmetry $\mathcal{H}_{\mathrm{K}'}=\mathcal{H}_{\mathrm{K}}^*$, and reads
%---------------------
\begin{equation}
  \mathcal{H}_{\mathrm{K}'}=\omega_{\text{D}}(L)\mathbb{1}-\mathrm{i}v_{\text{D}}(L)\boldsymbol{\sigma}^*\cdot\grad+\Phi(\mathbf{r},L)\mathbb{1}+\boldsymbol{\sigma}^*\cdot\mathbf{A}(\mathbf{r},L)\,,
  \label{eq:EffectHamiltonianKprime}
\end{equation}
%---------------------
where $\boldsymbol{\sigma}^*=(\sigma_x,-\sigma_y)$. Note, since time-reversal symmetry is preserved, the pseudo-vector potential, and the corresponding pseudo-magnetic field, couple with opposite signs in the two valleys. In equations\,\eqref{eq:EffectHamiltonianK} and \eqref{eq:EffectHamiltonianKprime} the Dirac frequency reads
%---------------------
\begin{equation}
\begin{split}
    \omega_{\text{D}}(L)=\;&\omega_0-\Omega\operatorname{Re}[\Sigma(\omega_0)]-3\Omega|\mathsf{G}_{\parallel}(\mathbf{a}_m)|\\
    &+\Omega\operatorname{Re}[\mathsf{G}^{\text{TEM}}_{\perp}(0,\omega_0)]-\sum_{\mathbf{G}}\frac{\Omega \omega_0^2\xi^2}{\omega_{\mathbf{K}-\mathbf{G}}^2-\omega_0^2}\,,
    \end{split}
    \label{eq:DiracFrequency}
\end{equation}
%---------------------
the Dirac velocity is
%---------------------
\begin{equation}
    v_{\text{D}}(L)=\frac{3\Omega a|\mathsf{G}_{\parallel}(\mathbf{e}_m)|}{2}-\sum_{\mathbf{G}}\frac{2\Omega \omega_0^2\xi^2 c^2 (\mathbf{K}-\mathbf{G})_x \phi_\mathbf{G}^{} }{(\omega_{\mathbf{K}-\mathbf{G}}^2-\omega_0^2)^2}\,,
    \label{eq:GroupVelocity}
\end{equation}
%---------------------
the strain-independent parameter in the pseudo-scalar potential reads
%---------------------
\begin{equation}
    \Phi_{0}(L)=\frac{3\Omega\beta_{\text{nnn}}|\mathsf{G}_{\parallel}(\mathbf{a}_m)|}{2}+\sum_{\mathbf{G}}\left[\frac{\Omega \omega_0^2\xi^2  }{\omega_{\mathbf{K}-\mathbf{G}}^2-\omega_0^2}-\frac{2\Omega \omega_0^2 \xi^2  c^2(\mathbf{K}-\mathbf{G})_x^2  }{(\omega_{\mathbf{K}-\mathbf{G}}^2-\omega_0^2)^2}\right]\,,
    \label{eq:PseudoScalarField}
\end{equation}
%---------------------
and, finally, the strain-independent parameter in the pseudo-vector potential is
%---------------------
\begin{equation}
    A_{0}(L)=\frac{3\Omega\beta_{\text{nn}}|\mathsf{G}_{\parallel}(\mathbf{e}_m)|}{4}-\sum_{\mathbf{G}}\frac{2\Omega \omega_0^2\xi^2c^2  (\mathbf{K}-\mathbf{G})_x^2\phi_\mathbf{G}^{}  }{(\omega_{\mathbf{K}-\mathbf{G}}^2-\omega_0^2)^2}\,.
    \label{eq:PseudoVectorField}
\end{equation}
%---------------------
Note,  to  evaluate  the  Dirac  frequency\,\eqref{eq:DiracFrequency}  one  requires  a  suitable  regularization  procedure since the last two terms separately diverge (see Supplementary Information for details).

%=============================================
\subsection{Numerical simulation of polariton wavepackets}
%=============================================

\noindent In this section we use the second-order split-operator method \cite{Chaves2010} to approximate the time evolution of the polariton envelope fields. After a small time $\delta t$ has elapsed, the polariton envelope field in the $\mathrm{K}$ valley is given by
%-------------------------------------
\begin{equation}
\psi_{\mathrm{K}}^{}(\mathbf{r},t+\delta t)=\mathrm{e}^{-\mathrm{i} \mathcal{H}_{\mathrm{K}}^{}\delta t}\psi_{\mathrm{K}}^{}(\mathbf{r},t)\,.
\label{eq:deltat}
\end{equation}
%-------------------------------------
In the second-order split operator method, the evolution operator in equation\,\eqref{eq:deltat} is approximated as 
%-------------------------------------
\begin{equation}
\mathrm{e}^{-\mathrm{i} \mathcal{H}_{\mathrm{K}}^{}\delta t}=\mathrm{e}^{-\frac{\mathrm{i}}{2}\mathcal{H}_{\mathrm{K}}^{\varepsilon}\delta t}\mathrm{e}^{-\mathrm{i}\mathcal{H}^{0}_{\mathrm{K}}\delta t}\mathrm{e}^{-\frac{\mathrm{i}}{2}\mathcal{H}_{\mathrm{K}}^{\varepsilon}\delta t}+\mathcal{O}(\delta t^3)\,,
\label{eq:EvolutionOperator}
\end{equation}
%-------------------------------------
where $\mathcal{H}^{0}_{\mathrm{K}}=\omega_{\text{D}}\mathbb{1}+\mathrm{i}v_{\text{D}}\boldsymbol{\sigma}\cdot\grad$ and $\mathcal{H}_{\mathrm{K}}^{\varepsilon}=\Phi(\mathbf{r})\mathbb{1}+\boldsymbol{\sigma}\cdot\mathbf{A}(\mathbf{r})$. Note, the cubic error in $\delta t$ is due to the noncommutativity of the position and gradient operators. To calculate the field after time $N_t \delta t$ has elapsed, we have to apply the operation\,\eqref{eq:EvolutionOperator} iteratively
%-------------------------------------
\begin{equation}
\psi_{\mathrm{K}}^{}(\mathbf{r},t+N_t\delta t)\approx \prod_{i=1}^{N_t}\left(\mathcal{M}_{r}^{\mathrm{K}}\mathcal{F}^{-1}\mathcal{M}_{k}^{\mathrm{K}}\mathcal{F}\mathcal{M}_{r}^{\mathrm{K}}\right)\psi_{\mathrm{K}}^{}(\mathbf{r},t)\,,
\label{eq:Iteration}
\end{equation}
%-------------------------------------
where $\mathcal{F}$ and $\mathcal{F}^{-1}$ represent the direct and inverse Fourier transform operations, respectively. Using the standard identity for the exponential of Pauli matrices, we can write the position-dependent operator in equation\,\eqref{eq:Iteration} as
%-------------------------------------
\begin{equation}
\mathcal{M}_{r}^{\mathrm{K}}=\mathrm{e}^{-\mathrm{i}\delta t\Phi/2}\left[\cos\big(\delta tA/2\big)\mathbb{1}-\mathrm{i}\frac{\sin\big(\delta t A/2\big)}{A}\boldsymbol{\sigma}\cdot \mathbf{A}\right]\,,
\label{eq:Mr}
\end{equation}
%-------------------------------------
and the momentum-dependent operator as
%-------------------------------------
\begin{equation}
\mathcal{M}_{k}^{\mathrm{K}}=\mathrm{e}^{-\mathrm{i} \omega_{\text{D}}^{}\delta t}\left[\cos\big(v_{\text{D}}k\delta t\big)\mathbb{1}+\mathrm{i}\frac{\sin\big(v_{\text{D}} k\delta t\big)}{k}\boldsymbol{\sigma}\cdot\mathbf{k}\right]\,.
\label{eq:Mk}
\end{equation}
%-------------------------------------
Similarly, the evolution of the polariton envelope field in the $\mathrm{K}'$ valley can be approximated as
%-------------------------------------
\begin{equation}
\psi_{\mathrm{K}'}^{}(\mathbf{r},t+N_t\delta t)\approx \prod_{i=1}^{N_t}\left(\mathcal{M}_{r}^{\mathrm{K}'}\mathcal{F}^{-1}\mathcal{M}_{k}^{\mathrm{K}'}\mathcal{F}\mathcal{M}_{r}^{\mathrm{K}'}\right)\psi_{\mathrm{K}'}^{}(\mathbf{r},t)\,,
\end{equation}
%---------------------------------------------------------------
where the operators $\mathcal{M}_{r}^{\mathrm{K}'}$ and $\mathcal{M}_{k}^{\mathrm{K}'}$ are related to equations\,\eqref{eq:Mr} and \eqref{eq:Mk} by the replacement $\boldsymbol{\sigma}\leftrightarrow\boldsymbol{\sigma}^*$ and $\mathbf{k}\leftrightarrow -\mathbf{k}$. 

For the simulations in figure\,\ref{fig:tunableorbits}c, we initialize the following Gaussian wavepackets
%-------------------------------------
\begin{equation}
\psi_{\mathrm{K}}^{}(\mathbf{r},t=0)=\frac{1}{2w\sqrt{2\pi}}
\mathrm{e}^{-\frac{|\mathbf{r}|^2}{2w^2}}\begin{pmatrix}
1 \\ -\operatorname{sign}(v_{\text{D}})
\end{pmatrix}\mathrm{e}^{\mathrm{i}\mathbf{k}_0\cdot \mathbf{r}}\,,
\end{equation}
%-------------------------------------
and
%-------------------------------------
\begin{equation}
\psi_{\mathrm{K}'}^{}(\mathbf{r},t=0)=\frac{1}{2w\sqrt{2\pi}}
\mathrm{e}^{-\frac{|\mathbf{r}|^2}{2w^2}}\begin{pmatrix}
1 \\ \operatorname{sign}(v_{\text{D}})
\end{pmatrix}\mathrm{e}^{\mathrm{i}\mathbf{k}_0\cdot \mathbf{r}}\,,
\end{equation}
%-------------------------------------
in the $\mathrm{K}$ and $\mathrm{K}'$ valleys, respectively. We consider wavepackets that are located in the lower polariton band with a fixed central frequency $\delta\omega=-0.001\omega_0$ relative to the Dirac point, and an initial central wavevector $\mathbf{k}_0=-|\delta\omega/v_{\text{D}}|\hat{\mathbf{x}}$ (i.e., initial group velocity in the $\hat{\mathbf{x}}$ direction). Furthermore, the wavepackets are initially centred at the origin with a width of $w=100a$. We then track the centre-of-mass trajectory of the wavepackets for the two valleys which is given by
%-------------------------------------
\begin{equation}
\langle\mathbf{r}\rangle_{\mathrm{K}/\mathrm{K}'}^{\vphantom{\dagger}}=\frac{\int \mathrm{d}^2\mathbf{r}\; |\psi_{\mathrm{K}/\mathrm{K}'}^{\vphantom{\dagger}}|^2\mathbf{r}}{\int \mathrm{d}^2\mathbf{r}\;|\psi_{\mathrm{K}/\mathrm{K}'}^{\vphantom{\dagger}}|^2}\,.
\end{equation}
%-------------------------------------
In the Supplementary information we go beyond the linear Dirac cone approximation by including second order field gradients in the effective Hamiltonian.

%=============================================
\subsection{Effective polarizability and local spectral function}
%=============================================

\noindent  In this section we go beyond the approximations of the effective Hamiltonian and derive the local spectral function which characterizes the full spectral response of the metasurface. To describe a scatterer's response to a local driving field, one has to take into account the strong multiple scattering within the metasurface. Therefore, we define an effective polarizability for a scatterer located at $\mathbf{r}_\alpha$ which is given by
%-------------------------------------
\begin{equation}
\alpha_{\text{eff}}^{-1}(\omega)=\alpha^{-1}(\omega)-\mathsf{S}(\mathbf{r}_\alpha,\mathbf{r}_\alpha,\omega )\,,
\end{equation}
%-------------------------------------
where the scattering function
%-------------------------------------
\begin{equation}
\mathsf{S}(\mathbf{r}_\alpha,\mathbf{r}_\alpha,\omega )=\sum_{\beta\neq\alpha}\sum_{\gamma\neq\alpha}  \mathsf{G}(\mathbf{r}_\alpha-\mathbf{r}_{\beta},\omega )[\bm{\mathsf{T}}^{(N)}(\omega)]_{\beta\gamma} \mathsf{G}(\mathbf{r}_{\gamma}-\mathbf{r}_\alpha,\omega )
\label{eq:scatterdyad}
\end{equation}
%-------------------------------------
encodes all the multiple-scattering events between the $N$ additional scatterers. In equation\,\eqref{eq:scatterdyad}, $\bm{\mathsf{T}}^{(N)}(\omega)$ is the $N$-scatterer T-matrix with matrix elements
%-------------------------------------
\begin{equation}
[\bm{\mathsf{T}}^{(N)}(\omega)]_{\beta\gamma}=\alpha(\omega)[ \bm{\mathsf{W}}^{-1}(\omega)]_{\beta\gamma}\,,
\end{equation}
%-------------------------------------
where the matrix elements of $\bm{\mathsf{W}}(\omega)$ read
%-------------------------------------
\begin{equation}
  [\bm{\mathsf{W}}(\omega)]_{\beta\gamma}=\delta_{\beta\gamma}-(1-\delta_{\beta\gamma})  \mathsf{G}(\mathbf{r}_{\beta}-\mathbf{r}_{\gamma},\omega ) \alpha(\omega)\,.
\end{equation}
%-------------------------------------
In figures\,\ref{fig:LDOSstrain}d-i, we plot the local spectral function $\operatorname{Im}[\alpha_{\text{eff}}(\omega)]$ for the scatterer located at $\mathbf{r}_{\alpha}=(0,a/2)$ on the $\mathrm{B}$ sublattice, and we include approximately 14,000 resonant dipole scatterers arranged in a circular configuration (before the applied strain).

\end{scriptsize}

%=============================================
%=============================================

%==================================================
%==================================================

\begin{footnotesize}

%=============================================
\subsection{Acknowledgments}
%=============================================

\noindent C.-R.M acknowledges financial support from the Engineering and Physical Sciences Research Council (EPSRC) of the United Kingdom through the EPSRC Centre for Doctoral Training in Metamaterials (Grant No. EP/L015331/1). S.A.R.H. acknowledges financial support from a Royal Society TATA University Research Fellowship (RPG-2016-186). E.M. acknowledges financial support from the Royal Society International Exchanges Grant IEC/R2/192166.

%=============================================
\subsection{Author contributions}
%=============================================

\noindent C.-R.M. conceived the idea, performed the theoretical calculations, and wrote the manuscript; S.A.R.H. contributed to the theoretical understanding; and E.M. contributed to the theoretical understanding and supervised the project; All authors commented on the manuscript.

%=============================================
\subsection{Additional information}
%=============================================

\noindent Supplementary information accompanies this work. Correspondence and requests for materials should be addressed to C.-R.M (email: cm433@exeter.ac.uk).

%=============================================
\subsection{Data availability}
%=============================================

\noindent All relevant data are available from the corresponding authors upon reasonable request.

%=============================================
\subsection{Competing interests}
%=============================================

\noindent The authors declare no competing interests.

\end{footnotesize}

%%TC:endignore
\clearpage
\end{document}